\documentclass[11pt]{article}
\begin{document}

\title{Cross-field transport and pitch-angle anisotropy of solar energetic particles in MHD turbulence}

\author{F. Fraschetti\footnote{Departments of Planetary Sciences and Astronomy, University of
Arizona, USA}}



\date{}
\maketitle

\begin{abstract}
Recent modelling of solar energetic particles (SEPs) propagation through the
heliospheric turbulence, also discussed in this workshop, has investigated
the role of the pitch-angle scattering and the perpendicular transport in
spreading particles in heliolongitude, as shown by multi-spacecraft
measurements (STEREO A/B, ACE, SOHO, etc.) at 1 AU in various energy
ranges. In some events the first-order pitch-angle anisotropy of the particles distribution is not-negligible. We calculate the average
perpendicular displacement due to the gradient/curvature drift in an
inhomogeneous turbulence accounting for pitch-angle dependence for two MHD turbulence models: (a) 3-D isotro\-pic, (b) anisotropic as conjectured by Goldreich-Sridhar. We find in both cases that the drift scales as $(1-\mu^2)^2$ with the cosine of pitch-angle $\mu$, in contrast with previous models for transport of SEPs. This
result can impact the models of propagation of SEPs through the heliosphere.
\end{abstract}

\section{Introduction}

In recent years, the multi-spacecraft monitoring by STEREO A/B, ACE, SOHO or
\textit{Ulysses}, has shown that SEP evciteents spread in longitude more than
$180^{\circ}$ at $1$ AU \cite{r14} and also in latitude as far as $70^{\circ}$
\cite{m09}. The interpretation of these measurements invokes two main
agents: (1) extended or moving source in the interplanetary medium (CME-driven
shocks); (2) strong perpendicular transport across the spiral magnetic field.
Measurements of first-order anisotropy suggest that the pitch-angle
dependence of the particle distribution is relevant to the interpretation of
the longitudinal spread \cite{d14}. {In situ} measurements of
longitudinal spread are particularly valuable in that they might probe
whether the transport perpendicular to the average magnetic field is
dominated by the meandering of the field lines or the transport away from the
field lines; this issue has been recently investigated in \cite{fj11}.

Test-particle simulations \cite{fg12} show that even in a weak
three-dimensional isotropic turbulence charged particles decorrelate from the
unperturbed magnetic field on a time scale comparable to the gyroperiod. In
addition to repeated scattering in the pitch-angle by which particles move
back and forth along a given field line, a prompt decorrelation reduces the
initial anisotropy of the particles streaming from the source and enhances
the angular spread. These results pinpoint to a propagation regime wherein
early-time perpendicular transport cannot be neglected.

Recent phenomenological models of charged particle transport in the
heliosphere include a dependence of the perpendicular diffusion coefficient
on the pitch-angle of the particle velocity with respect to the global
average magnetic field ($\mu$) given by $\sqrt{1-\mu^2}$, suggested by a
proportionality of the perpendicular diffusion coefficient to the particle
gyroradius \cite{d14}. Different pitch-angle dependencies were considered in
\cite{sf15}.

In this short note we outline the calculation of the perpendicular transport
coefficient due to gradient/curvature drift originating from the
inhomogeneity of the magnetic turbulence for an anisotropic pitch-angle
distribution of the particles. The results for the 3-D isotropic turbulence
and the MHD anisotropic turbulence conjectured by \cite{gs95}, hereafter
GS95, are compared. In \cite{fj11} we have calculated the average over an
isotropic pitch-angle distribution; in this short note we relax such an
assumption. We do not calculate here the contribution to the perpendicular
transport due to field line meandering.

\section{Gradient/curvature drift transport}\label{sect_drift}

\subsection{Three-dimensional isotropic}

We consider a spatially homogeneous
time-independent magnetic field with superimposed fluctuations. The amplitude
of the fluctuation (${\delta B}$) is assumed to be much smaller than the
average field magnitude ($B_0$). We represent such a magnetic field as
${\mathbf{B}}(\mathbf{x}) =\mathbf{B}_0 + \delta {\mathbf{B}(\mathbf{x})}$, with an average component $\mathbf{B}_0 = B_0
\mathbf{e}_z$; we assume that the average over field realisations vanishes,
$\langle \delta \mathbf{B}(\mathbf{x}) \rangle = 0$, and $\delta B(\mathbf{x})/B_0 \ll 1$.
We make use of the gyroperiod averaged guiding-center velocity transverse to
the field $\mathbf{B} (\mathbf{x})$, i.e., $\mathbf{V}_{\perp}^G (t)$, to the first order in
$\delta B(\mathbf{x})/B_0$, given for a particle of speed $v$, momentum $p$ and
charge $Ze$ by
\begin{equation}
\mathbf{V}_{\perp}^G (t) =  \frac{vpc}{Ze B} \left[ \frac{1-\mu^2}{2}
\frac{\mathbf{B}\times \nabla B}{B^2} + \mu^2 \frac{\mathbf{B} \times (\mathbf{B} \cdot \nabla ) \mathbf{B})}{B^3}   \right]
\label{Vperp}
\end{equation}
where $\mu$ is the cosine of the pitch-angle with respect to the average
field $B_0$ \cite{ro70,fj11}. The average square transverse displacement of
the particle guiding center from the direction of local $B$ due to drift
along the axis $x_i$, $d_{D_{ii}} (t)$, at time $t$ is written as
\begin{equation}
d_{D_{ii}} (t) = \int _0 ^t  d\xi \langle {\bf V}_{\perp, i}^G (t') {\bf V}_{\perp, i}^G (t' + \xi) \rangle  .
\label{dXX1}
\end{equation}

In the Eq.~(\ref{Vperp}) we make use of the Fourier representation for $\delta
\mathbf{B}(\mathbf{x})$ and assume an inertial range magnetic turbulence power
spectrum which is uncorrelated at different wavenumber vectors: $\langle
\delta B_r({\bf k}) \delta B_q ^*(\mathbf{k}') \rangle = \delta (\mathbf{k} - \mathbf{k}')
P_{rq}(\mathbf{k})$. As in \cite{fj11} we use
\begin{equation}
 P_{rq} ({\bf k}) = \frac{G(k)}{8\pi k^2} \left[ \delta_{lm} - \frac{k_l k_m}{k^2}\right], \\
\end{equation}
 with
\begin{equation} 
 G(k)= \left\{
  \begin{array}{cc}
    G_0 k ^{-q} & \rm{if} k_{\it min} < k  <  k_{\it max}  \, \\
    G_0 k_{\it min} ^{-q} & \rm{if} k_0< k  <  k_{\it min}\, ,
   \end{array}
\right.
\end{equation}
where $k^{\it max}$ corresponds to the scale where the dissipation rates of the
turbulence overcomes the energy cascade rate, the coherence length is given
by $L = 2\pi/k^{\it min}$, and the physical scale of the system by $2\pi/k ^0$.

With these assumptions, the cross-field diffusion coefficient (limit for large times of the average square displacement) associated to gradient/curvature
drift of the guiding center due to the turbulence inhomogeneity (in other
words the diffusive motion of the guiding center away from the field lines)
reads
\begin{equation}
D^i _D (t) \rightarrow   { \pi\over 16} \left( \frac{\delta B}{B_0} \right)^2 \left
 | \frac{q-1}{q-2} \right | \frac{2\pi v^3}{L\Omega^2} \left ( \frac{1 - \mu^2}{2}\right)^2
\label{dDiso_asy}
\end{equation}

We note that $1/2 \int_{-1}^{1} d\mu D^i _D (t) = d^i _D (t)$, where $d^i _D
(t)$ is the time-dependent average square displacement in Eq.~(45) of
\cite{fj11}. The condition that the perpendicular transport time-scale is
much shorter than the parallel one is translated into the restriction of Eq.~(\ref{dDiso_asy})
to times $t < 1/k_{\it min} v_\parallel \simeq L/ 2 \pi
v_\parallel$; thus, for higher energy particles the physical time this regime
applies is shorter because the perpendicular scale grows with the particle
gyroradius $r_\mathrm{g}$ and the perpendicular transport becomes relevant. For a $25$
MeV proton (within the energy range of STEREOs and SOHO instruments) at $1$
AU, $(\delta B/B_0)^2 = 0.1$, $r_\mathrm{g}/L \simeq 0.1$ ($L = 0.01$~AU) and $q =
11/3$, we have $D^i_{D} (t) \simeq 4.5 \times 10^{17} (1 - \mu^2)^2 $\,cm$^2$\,s$^{-1}$
for $t < 1/k_{\it min} v_\parallel \simeq 3.5$\,s.

\subsection{MHD anisotropic turbulence GS95}
Also in this case we consider a
spatially homogeneous, fluctuating, time-independent global magnetic field
$\mathbf{B}_0$ and decompose the field as ${\mathbf{B}}(\mathbf{x}) =\mathbf{B}_0 + \delta {\mathbf{B}(\mathbf{x})}$,
with a large-scale average component $\mathbf{B}_0 = B_0 \mathbf{e}_z$ and
a fluctuation $\langle \delta \mathbf{B}(\mathbf{x}) \rangle = 0$. We assume that the
inertial range extends from $L$ down to some scale wherein injected energy
ultimately must dissipate. We introduce a scale $\hat L$ such that $r_\mathrm{g} \ll
\hat L \ll L$. The fluctuations are large compared to the total magnetic
field at scale $\sim L$; however, at scales $< \hat L$ the fluctuations are
small compared to the local average field, and the first-order orbit theory
applies to eddies up to scale $\hat L$. By using similar assumptions,
\cite{c00} made use of the quasi-linear theory to calculate parallel
transport coefficient in GS95.

According to the GS95 conjecture, the pseudo-Alfv\'en modes are carried
passively by the shear-Alfv\'en modes with no contribution to the turbulence
cascade to small scales which is seeded by collisions of shear modes only.
{\it \cite{f15}} shows that the transport perpendicular to the local average field
due to gradient/curvature drift is dominated by the power of the
pseudo-Alfv\'en modes along the local average field given by
\begin{equation}
P_{33} (\hat k_\parallel, \hat k_\perp) = \frac{\hat k_\perp ^2}{\hat k^2}
\Pi(\hat k_\parallel, \hat k_\perp),   \\
\end{equation}
with 
\begin{equation}
{ \Pi
(\hat k_\parallel, \hat  k_\perp)  =  \frac{\varepsilon {\cal N} B_0^2}{ \ell^{1/3}}
 \hat k_\perp^{-10/3} {\exp}\left( -\frac{\ell^{1/3} \hat  k_\parallel}{\hat k_\perp^{2/3}}\right)} .
\label{pseudo}
\end{equation}
where $\hat k_\parallel, \hat k_\perp$ indicate wave numbers parallel and
perpendicular, respectively, to the local average field, $\ell$ is the outer
scale of the pseudo-modes, $\varepsilon$ is the power in the pseudo- relative
to shear-modes ($0 < \varepsilon < 1$) and ${\cal N}$ is a normalisation constant
accounting for both polarisation modes given by ${\cal N} \simeq {1\over 3\pi} (\delta B/B_0)^2
(1 + \varepsilon (L/\ell))^{-1}$, that includes also scales between $\hat L$ and $L$, as it
can be easily seen that the contribution to the turbulent power from perpendicular
scales $\hat L < k_\perp ^{-1} < L$ is exponentially suppressed.

Following the calculation in the Appendix of \cite{f15}, we find that the average square displacement due to gradient/curvature drift is given by
\begin{equation}
 D^a_{D} (t) \simeq \frac{1}{8} \left(\frac{\delta B}{B_0} \right)^2 \frac{\varepsilon}{1+\varepsilon L/\ell}
  \left(\frac{L}{\ell} \right)^{1/3}   \frac{v^4}{\Omega ^3 L^2}  {\Omega t} (k_\perp ^M L)^{2.2} \left ( \frac{1 - \mu^2}{2}\right)^2 .
\label{series_tot_drift}
\end{equation}

Due to the lack of space, we omit here the corresponding result for the
shear- modes. For a 25\,MeV proton at 1~AU, by taking $L = l$, $\varepsilon =
0.5$, $(\delta B/B_0)^2 = 0.1$ and $(k_\perp ^M)^{-1} \simeq 10^{-5}$~AU, we
have $D^a_{D} (t) \simeq 2.5 \times 10^{21} (1 - \mu^2)^2 t [{\it s}] $ cm$^2$\,s$^{-1}$
for $t < 1/k_{\parallel}^{\it min} v_\parallel \simeq 3.5$\,s. We
emphasise that such a drift cumulates for hours, that is the typical
time-scale of the longitudinal spread of SEP, resulting in a possibly
significant contribution.

\section{Conclusions}

We have calculated the time-dependent average square displacement DD(t) due
to gradient/curvature drift for two distinct MHD turbulence models: (a)
3-D-isotropic and (b) anisotropic as conjectured by Goldreich and
Sridhar~(1995); the assumption of the pitch-angle isotropy of the particle
distribution function has been relaxed to account for recently measured first
order anisotropy. In both cases, we find $D_D \propto (1 - \mu^2)^2$; such a
dependence arises from the gradient drift that is proportional to the
particle kinetic energy normal to the field ($p_\perp ^2 / 2m$ for a particle
of mass $m$). We conclude that spacecraft data compatible with the scaling
$(1 - \mu^2)^2$ would support at once the Goldreich-Sridhar conjecture and
our model for perpendicular transport. Although drifts cannot be neglected in
the interpretation of multi-spacecraft SEP data across the Parker spiral
\cite{m13}, the meandering of the field lines is expected to be larger than
drifts in perpendicular transport at scales close to the outer scale. This
effect will be assessed in a separate work.

\section{Acknowledgements}
The author acknowledges useful discussions with W. Dr\"oge and R. D. Strauss
and constructive feedback of the referees. This work was supported, in part,
by NASA under grant NNX13AG10G. This work benefited from discussions at the
team meetings ``First principles physics for charged particle transport in
strong space and astrophysical magnetic turbulence'' at ISSI in Bern,
Switzerland.



\begin{thebibliography}{}

\bibitem{c00}
Chandran, B. D. G.: Scattering of Energetic Particles by Anisotropic
Magnetohydrodynamic Turbulence with a Goldreich-Sridhar Power Spectrum,  Phys. Rev. Lett., 85, 4656--4659,
2000.

\bibitem{d14}
Dr\"oge, W., Kartavykh, Y. Y., Dresing, N., Heber, B., and Klassen, A.:
Wide longitudinal distribution of interplanetary electrons following the
7 February 2010 solar event: Observations and transport modeling,  J. Geophys. Res., 119, 6074--6094,
2014.


\bibitem{f15}
Fraschetti, F.: Cross-field transport in Goldreich-Sridhar MHD turbulence
2015, pre-print: arXiv/1512.05352.

\bibitem{fg12}
Fraschetti, F. and Giacalone, J.: Early-time velocity auto-correlation for
charged particles diffusion and drift in static magnetic turbulence,  Astrophys. J., 755, 114, 9 pp.,
2012.

\bibitem{fj11}
Fraschetti, F. and  Jokipii, J. R.:  Time-dependent perpendicular transport
of fast charged particles in a turbulent magnetic field, Astrophys. J., 734, 83, 8 pp.,
2011.


\bibitem{gs95}
Goldreich, P. and Sridhar, S.: Toward a theory of interstellar turbulence.
2: Strong alfvenic turbulence,  Astrophys. J., 438, 763--775, 1995.

\bibitem{m09}
Malandraki, O. E., Marsden, R. G., Lario, D., Tranquille, C., Heber, B., Mewaldt,
R. A., Cohen, C. M. S., Lanzerotti, L. J., Forsyth, R. J., Elliott, H. A., Vogiatzis, I. I., and Geranios, A.:
Energetic Particle Observations and Propagation in the Three-dimensional Heliosphere During the 2006 December Events,  Astrophys. J., 704, 469--476,
2009.

\bibitem{m13}
Marsh, M. S., Dalla, S., Kelly, J., and Laitinen, T.: Drift-induced
Perpendicular Transport of Solar Energetic Particles,  Astrophys. J., 774, 4, 9 pp.,
2013.

\bibitem{r14}
Richardson, I. G., von Rosenvinge, T. T., Cane, H. V., Christian, E. R., Cohen,
C. M. S., Labrador, A. W., Leske, R. A., Mewaldt, R. A., Wiedenbeck, M. E., and Stone,
E. C.: $>$\,25\,MeV Proton Events Observed by the High Energy Telescopes on the
STEREO A and B Spacecraft and/or at Earth During the First Seven Years of the STEREO Mission, Sol. Phys., 289, 3059--3107,
2014.

\bibitem{ro70}
Rossi, B. and Olbert, S.: Introduction to the Physics of Space,  McGraw-Hill,
New York, 1970.

\bibitem{sf15}
Strauss, R. D. and  Fichtner, H.: On aspects pertaining to the perpendicular
diffusion of solar energetic particles,  Astrophys. J., 801, 29, 2015.

\end{thebibliography}
\end{document}